\documentclass[twocolumn,showpacs,preprintnumbers,amsmath,amssymb]{revtex4-1}

\usepackage{graphicx}
\usepackage{amssymb}
\usepackage{hyperref}

\begin{document}

\title{Emerging frustration effects in ferromagnetic
Ce$_2$(Pd$_{1-x}$Ag$_x$)$_2$In alloys}

\author{J.G. Sereni $^1$, M. Giovannini $^{2,3}$, M. G\'omez Berisso$^{1}$, F. Gastaldo$^{2}$}
\address{$^1$ Low Temperature Div. CAB - CNEA, Conicet, 8400 Bariloche, Argentina}
\address{$^2$ Dip. di Chimica e Chimica Industriale, Universit$\grave{a}$ di Genova, I-16146 Genova,
Italy}
\address{$^3$ CNR-SPIN Corso Perrone 16152 Genova, Italy}

\begin{abstract}

{Magnetic and thermal properties of Ferromagnetic (FM)
Ce$_{2.15}$(Pd$_{1-x}$Ag$_x$)$_{1.95}$In$_{0.9}$ alloys were
studied in order to determine the Quantum Critical Point (QCP) at
$T_C \to 0$. The increase of band electrons produced by Pd/Ag
substitution depresses $T_C(x)$ from 4.1\,K down to
$T_C(x=0.5)$=1.1\,K, with a QCP extrapolated to $x_{QCP} \geq
0.5$. Magnetic susceptibility from $T>30$\,K indicates an
effective moment slightly decreasing from $\mu_{eff}$=2.56$\mu_B$
to $2.4\mu_B$ at $x$=0.5. These values and the paramagnetic
temperature $\theta_P\approx$ -10\,K exclude significant Kondo
screening effects. The $T_C(x)$ reduction is accompanied by a
weakening of the FM magnetization and the emergence of a specific
heat $C_m(T)$ anomaly at $T^*\approx 1$\,K, without signs of
magnetism detected from AC-susceptibility. The magnetic entropy
collected around 4\,K (i.e. the $T_C$ of the $x=0$ sample)
practically does not change with Ag concentration:
$S_m(4K)\approx$ 0.8 Rln2, suggesting a progressive transfer of FM
degrees of freedom to the non-magnetic (NM) component. No
antecedent was found concerning any NM anomaly emerging from a FM
system at such temperature. The origin of this anomaly is
attributed to an {\it entropy bottleneck} originated in the nearly
divergent power law dependence for $T>T^*$.}
\end{abstract}

\date{\today}

\maketitle


\section{Introduction}
\label{Introduction}

The R$_2$T$_2$X family of compounds (with R = Rare earth, T =
transition metal and X = semi metal) were actively investigated
during the last decade because of their peculiar magnetic
properties at low temperatures (see e.g. \cite{Muramatsu}). The
strongly anisotropic Mo$_2$B$_2$Fe type crystalline structure
\cite{Fourgeot96}, with alternated magnetic and non-magnetic
atomic layers, favors geometrical frustration effects originated
in the triangular coordination of the R-magnetic atoms.

\begin{figure}[tb]
\begin{center}
\includegraphics[width=19pc]{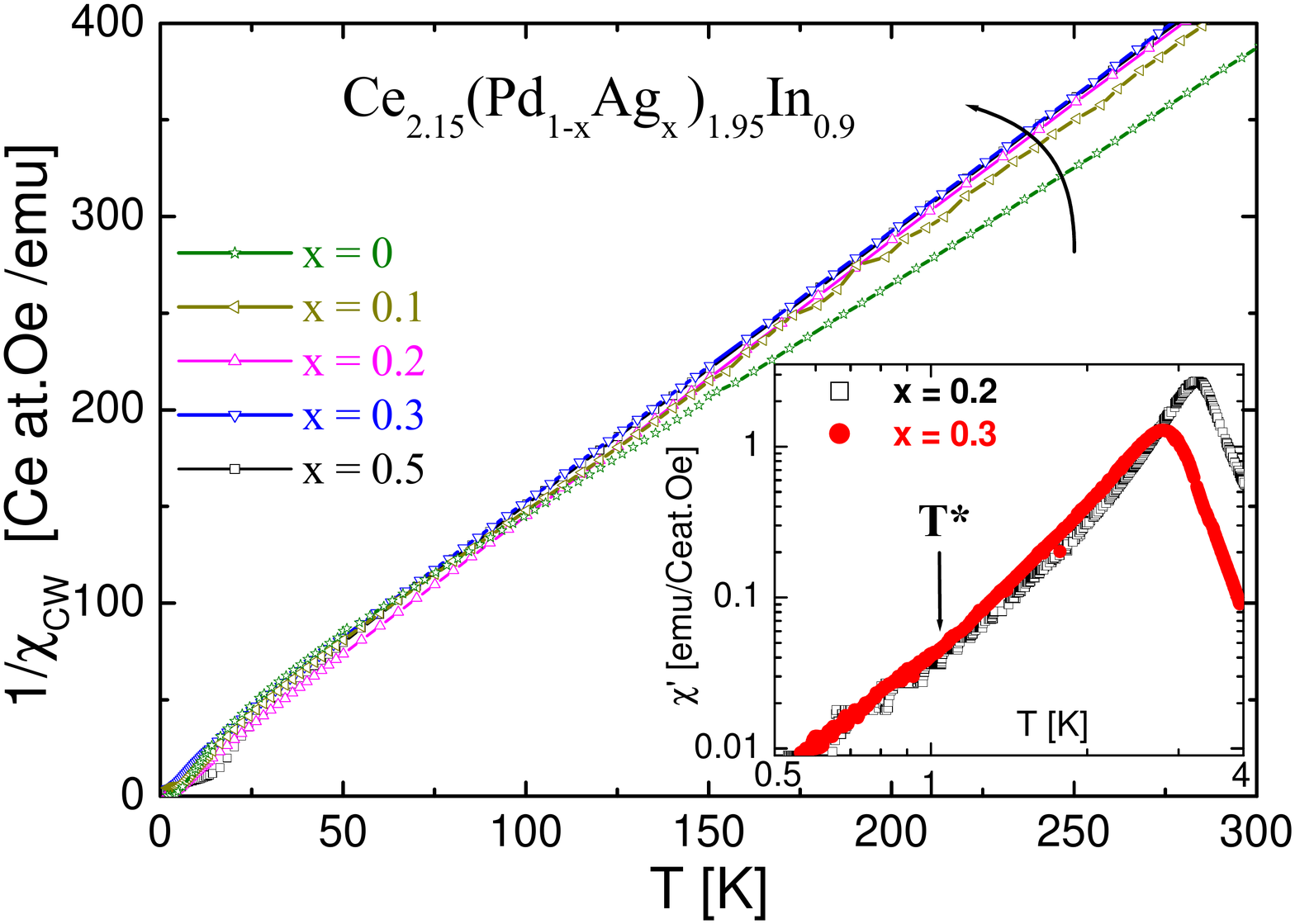}
\end{center}
\caption{(Color online) Inverse of the Curie-Weiss contribution of
magnetic susceptibility, after the Pauli-like contribution
subtraction, measured up to room temperature in a field of
0.5\,Tesla. The arrow indicates the increase of Ag concentration.
Inset: Monotonous decrease of the inductive component of
AC-susceptibility ($\chi'$) for samples $x=0.2$ and 0.3 on the FM
phase ($T<T_C$) in a double logarithmic representation. T$^*$
indicates the temperature of the specific heat anomaly growing
with Ag content, see the text.} \label{F1}
\end{figure}

The extended range of solid solution of the Ce$_{2\pm u}$Pd$_{2\mp
y}$In$_{1-z}$ system \cite{Mauro00} has allowed to determine that
Ferro (FM) or Antiferromagnetic (AFM) behaviors depend on the
relative concentration of {\it electron-holes} in the T-X layer
\cite{SoliSolut}. In fact, the Ce-rich (i.e. electron rich) branch
behaves FM, whereas the Pd-rich (i.e. hole rich) behaves AFM. In
this crystalline structure the triangular coordination of next
magnetic neighbors ($nmn$) fulfils the condition for magnetic
frustration within Ce planes \cite{Ramirez}, provided there are
AFM interactions between Ce-$nmn$ within the plane.

Exploiting the fact that a FM ground state can be driven by tuning
the electron-hole concentration, the
Ce$_{2.15}$Pd$_{1.95}$In$_{0.9}$ composition was chosen as a
starting point to approach a FM quantum critical regime by doping
Pd lattice with Ag in
Ce$_{2.15}$(Pd$_{1-x}$Ag$_x$)$_{1.95}$In$_{0.9}$ alloys. The
starting concentration $x=0$ lies in the vicinity of a magnetic
critical point, determined in a previous investigation performed
with Rh doping by the presence of two magnetic transitions
converging to a critical point at the vicinity of $x=0$
\cite{SCES_PdRhIn}.

\begin{figure}[tb]
\begin{center}
\includegraphics[width=19pc]{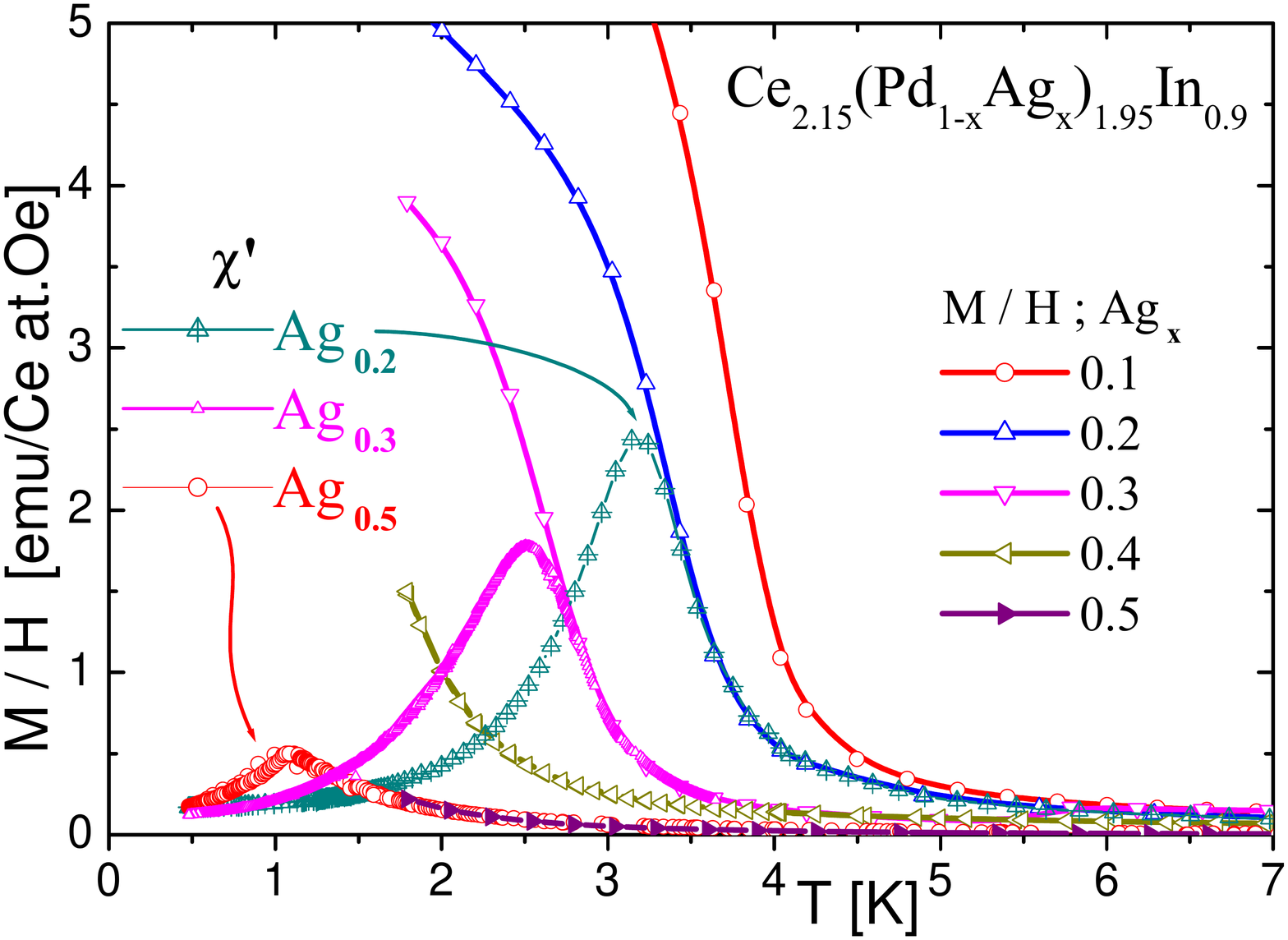}
\end{center}
\caption{(Color online) Low temperature dependence of the
magnetization measured with $H=0.1$\,Tesla showing the FM
character of the system. The inductive component of the
AC-susceptibility ($\chi'$) of some representative samples are
scaled to $M/H$ units of respective concentrations using a unique
scaling factor. The $\chi'$ maxima mark the $T_C(x)$ temperatures
that show a progressive decrease of with Ag content.} \label{F2}
\end{figure}

\section{Experimental results}

\subsection{Sample preparation and Characterization}

The samples were prepared using a standard arc melting procedure
under an argon atmosphere, and they were remelted several times to
ensure good homogeneity. The
Ce$_{2.15}$(Pd$_{1-x}$Ag$_x$)$_{1.95}$In$_{0.9}$ alloys form
continuously up to the limit of solubility at $x=0.5$ within the
Mo$_2$B$_2$Fe-type structure. The volume of the unit cell
increases with Ag content following a Vegard's law up to $x=0.5$,
with the 'c/a' ratio remaining nearly constant.

\subsection{Magnetic Properties}

High temperature ($T>30$\,K) magnetic susceptibility results are
properly described by a $\chi=\chi_{cw}+\chi_p$ dependence, where
the first term corresponds to the temperature dependent
Curie-Weiss contribution $\chi_{cw}= \frac{Cc}{T+\theta}$ and the
second to a Pauli-like contribution. This $\chi_p$ contribution is
observed along the full concentration range with a value of
$\chi_p=(2\pm 0.5) 10^{-4}$\,emu/molCeOe. From the inverse of
$\chi_{cw}$ (see Fig.~\ref{F1}) one extracts the Curie constant
(Cc) which indicates the full development of a Ce$^{3+}$ magnetic
moment (i.e. $\mu_{eff} = 2.56 \mu_B$ per Ce atom) for the pure Pd
alloy. This value slightly decreases down to $\approx 2.4 \mu_B$
at $x=0.5$. The paramagnetic temperature $\theta_P$ practically
does not change with concentration, remaining around $\theta_P
\approx -10$\,K. This negative value is evaluated as an
extrapolation of $1/\chi_{cw}$ from $T>30$\,K down to
$1/\chi_{cw}=0$. At this temperature range the crystal electric
field states contribute significantly. Below $T\approx 30$\,K, a
moderate downward curvature makes $1/\chi_{cw}$ to extrapolate to
$T>0$ revealing the FM character of the ground state.

Since $\chi_{cw}(T)$ measurements are limited down to $T=1.8$\,K,
we have extended the study of the magnetic properties performing
AC-susceptibility ($\chi'$) measurements down to 0.5\,K on some
representative samples. In the inset of Fig.~\ref{F1}, the
$\chi'(T<4\,K)$ results from samples $x =0.2$ and 0.3 are
presented in a double logarithmic representation that covers more
than two decades of magnetic signal intensity.

In Fig.~\ref{F2}, the low temperature dependence of the
magnetization $M(T)$, measured at $H=0.1$\,Tesla, is presented as
a $M/H$ ratio. The upturn at $T\leq 4$\,K reveals the FM character
of the ground state (GS). Notably, the measured magnetization
decreases hand in hand with $T_C(x)$ decrease. The inductive
component $\chi'$ of the measured samples is also included in this
figure in order to compare them with $M/H$ measurements. One can
observe the coincident decrease of the $\chi'$ signal with
$T_C(x)$, the latter identified by the maximum of $\chi'(T)$ and
the maximum slope of $\partial M/\partial T$. The $\chi'(T)$
results from sample $x=0.5$ are also included in Fig.~\ref{F2} to
obtain the $T_C(x=0.5)=1$\,K value and to confirm the continuous
vanishing of the FM signal.

Magnetization $M(H)$ curves at $T=1.8$\,K (not shown) of samples
$x= 0.1$ and 0.2 reach $90\%$ of its saturation value $M_{sat} =
1.1\mu_B$/Ce\,at. at $H\approx 0.5$\,Tesla, showing a small
hysteresis loop, typical for FM materials. Above that
concentration ($x\geq 0.3$), the initial $M(H)$ slope weakens and
transforms into a continuous curvature. In spite of that, the
$x=0.5$ alloy still shows a remanent loop of hysteresis from the
vanishing FM component between $-0.2 < H < 0.2$\,Tesla and $-0.15
< M < 0.15 \mu_B$/Ce\,at. This small contribution has to be
compared with the total magnetization reached at $H=5$\,Tesla:
$0.9\mu_B$/Ce\,at.

\begin{figure}[tb]
\begin{center}
\includegraphics[width=19pc]{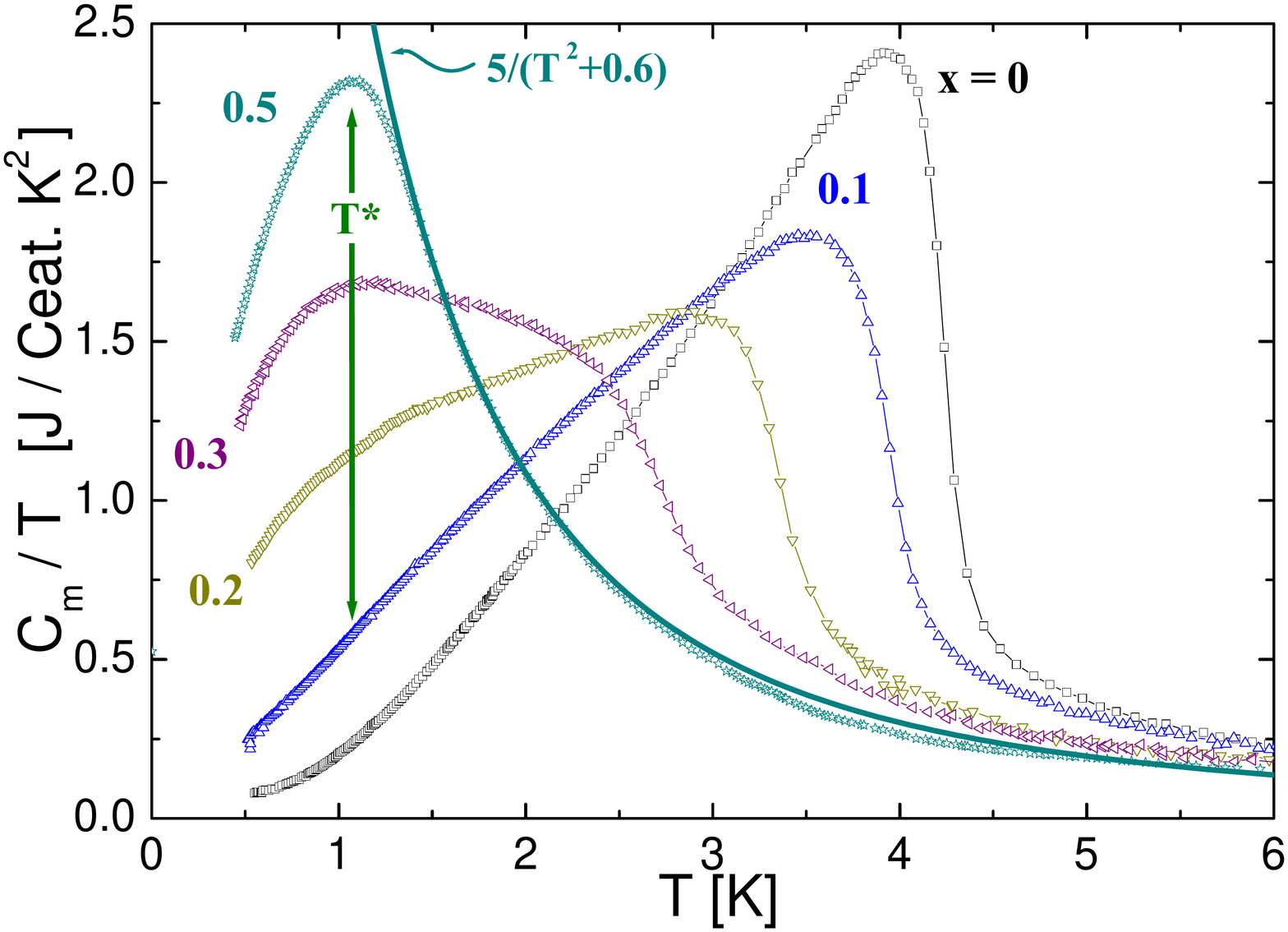}
\end{center}
\caption{(Color online) Specific heat of
Ce$_{2.15}$(Pd$_{1-x}$Ag$_x$)$_{1.95}$In$_{0.9}$ alloys. The
continuous curve is the fit of $C_m/T$ for $T>T^*$ in sample
$x=0.5$. Green arrow indicates the emerging anomaly at fixed
temperature $T=T^*$} \label{F3}
\end{figure}

\subsection{Specific heat}

The $T_C(x)$ transition is recognized in specific heat ($C_m$)
measurements by a clear jump $\Delta C_m$ which, for the mother
compound Ce$_{2.15}$Pd$_{1.95}$In$_{0.9}$, reaches the value of
$\Delta C_m \approx 10$\,J/Ceat.K (see Fig.~\ref{F3}). This value
is close to the value predicted for a doublet GS \cite{Morphol}.
At very low temperature, a $C_m(T)$ curvature can be fitted with
an exponential function indicating the presence of a gap in the
magnon spectrum typically occurring in strongly anisotropic
systems. The magnetic contribution to the specific heat $C_m$ is
obtained after subtracting the phonon contribution extracted from
a La$_{2}$Pd$_2$In compound.

The $\Delta C_m/T$ jump progressively decreases and broadens as
$T_C(x)$ decreases. Unexpectedly, another anomaly emerges around
$T^* \approx 1$\,K, overcoming the FM $\Delta C_m$ jump around
$x=0.3$, but without changing its position in temperature with Ag
content. This anomaly seems to be fully developed for the $x=0.5$
alloy. It cannot be associated neither to spin glass ($C_{sg}$)
nor to Schottky ($C_{sh}$) anomalies because those specific heat
anomalies have an associated magnetic signal and because they do
not describe the observed $C_m/T$ thermal dependence on both sides
of the maximum properly. In fact, below the respective maxima,
spin glasses show a $C_{sg}/T = const.$ behavior \cite{Mydosh} and
$C_{sh}\propto \exp(1/T)$ \cite{Tari}, compared with the
$C_m/T\propto T$ of sample $x=0.5$. Above the maximum both
specific heat anomalies decay as $C_p/T \propto 1/T^3$, which is
different from the temperature dependence observed. In the
following section we discuss a tentative description of the
$C_m(T)/T$ tail at $T>T^*$ using a modified power law $C_m(T)/T$
dependence together with the thermodynamical implications of such
a thermal dependence. Notably, the total entropy gain evaluated as
$S_m(T)= \int C_m(T)/T\,dT$ up to $T=7$\,K practically does not
change with Ag content, and reaches $\approx 90\%$ of R$\ln2$ per
Ce atom at T=10\,K.

\section{Discussion}

Despite the fact that high temperature susceptibility measurements
reveals a robust magnetic moment in Ce ions only weakly dependent
on Ag concentration, $T_C(x)$ decreases hand in hand with the
intensity of the FM signal. This evolution cannot be explained by
usual Kondo screening of Ce magnetic moments because $-\theta_P(x)
\propto T_K$ slightly depends on concentration. The increase of
Ce-$nmn$ spacing, driven by the expansion of the lattice
parameters with Ag content, can be only partially responsible for
the weakening of the RKKY interaction because it increases about
$1.4\%$ between x=0 and 0.5. Similar expansion occurs between Ce
planes reflected in the increase of the 'c' axis.

The outstanding message from the $\chi'(T)$ dependence measured on
samples $x=0.2$ and 0.3, presented in the inset of Fig.~\ref{F1},
is the decrease of the magnetic response below $T_C$ without any
detected magnetic contribution around $T^* \approx 1$\,K. The
double logarithmic representation, covering more than two decades
of signal variation, shows a monotonous behavior that excludes
other contributions. Notice that in the alloy with $x = 0.3$ the
$T^*$ anomaly already contains a similar amount of degrees of
freedom like the decreasing FM component as it can be appreciated
from $C_m/T$ measurements in Fig.~\ref{F3}. In Fig.~\ref{F2} these
two $\chi'(T)$ curves and the one from $x=0.5$ are scaled to
respective $\chi_{cw}$ values above 1.8\,K using a unique scaling
factor between the induced AC-voltage and the magnetic units. From
this comparison one can appreciate how the maximum of the
$\chi'(T)$ signal at $T=T_C$ decreases together with $T_C(x)$,
extrapolating the critical concentration $x_{cr}$ slightly beyond
0.5.

\begin{figure}[tb]
\begin{center}
\includegraphics[width=19pc]{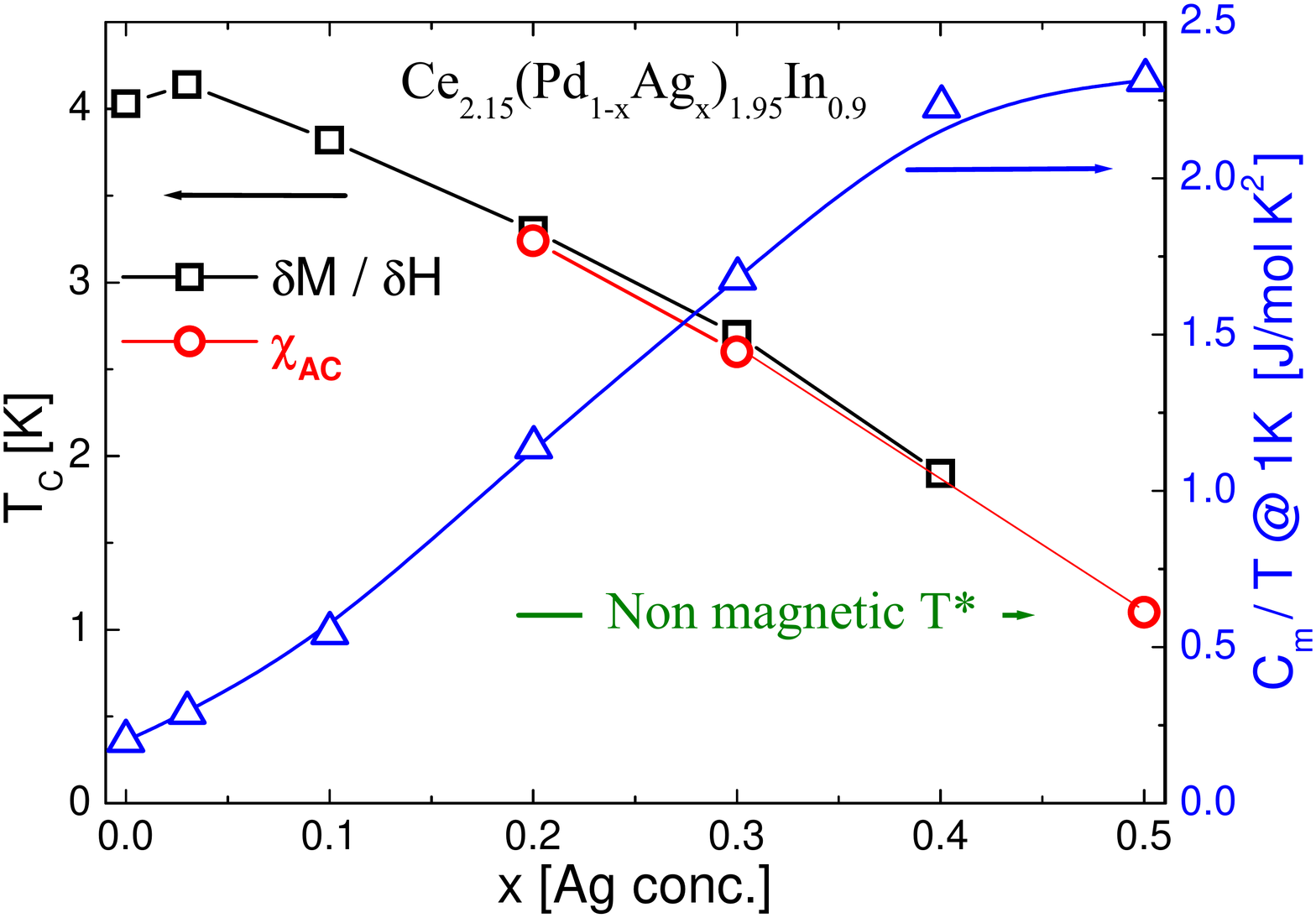}
\end{center}
\caption{(Color online) Left axis: Concentration dependence of
$T_C(x)$ extracted from the maximum slope of $M(T)$ for $x\leq
0.4$ alloys (black squares) and from the maximum of $\chi'$ for
$x\geq 0.2$ (red circles). Right axis: $C_m/T$ value at $T=T^*$.
The Non magnetic $T^*$ arrow indicates the temperature of the
maximum of the Non magnetic anomaly at $T\approx 1$\,K.}
\label{F4}
\end{figure}

Additionally to the transference of degrees of freedom from FM
component to a non-magnetic component observed in the $C_m(T,x)/T$
dependence, presented in Fig.~\ref{F3}, a more quantitative
indication for such transference can be appreciated in
Fig.~\ref{F4} where the value of $C_m(x)/T$ at $T=T^*$ is
depicted. From the figure it can be seen that there is a clear
increase of $C_m/T^*$ (right axis) as the FM component vanishes
proportionally to $T_C$ decrease (left axis). The same analysis
can be done in terms of the entropy accumulated within the
anomaly. However, at intermediate concentrations, the subtraction
of each component from the total $C_m/T$ value becomes difficult.

The unexpected absence of magnetic signal from the $T^*$ anomaly
is the outstanding feature of this system. To our knowledge there
is no antecedent reported in the literature for a {\it
non-magnetic} anomaly at such a low temperature, at least for Ce
compounds. Although some degree of magnetic disorder can be
expected in the Pd/Ag plane due to the different atomic size of
the atoms, this effect usually drives the magnetic system into a
concentration dependent spin glass type behavior, not observed in
this $C_m(T)$ anomaly. Furthermore, there is no evidence that
magnetic disorder weakens a FM interaction between neighbors. For
comparison, we observe that under magnetic field this transference
of degrees of freedom is reversed respect to the Ag concentration
effect, weakening the $C_m(T^*)$ anomaly but without detecting a
change in its temperature position. Since the non-magnetic degrees
of freedom are related to magnetically frustrated moments,
magnetic field shall reduce frustration because it favors FM
alignment.

The relevant questions arising about this system are: the origin,
its non magnetic nature and the low characteristic temperature
$T^*=1$\,K. The origin of this anomaly may be attributed to
frustration effects because a change of sign in the RKKY magnetic
interaction can be expected. Since this interaction is mediated by
conduction electrons, the change from 'd'-hole Pd to 's'-electron
Ag ligands modifies the polarization of the exchange interaction.
Then, if an AF interaction sets on, in the Mo$_2$B$_2$Fe type
structure the triangular configuration between Ce-$nmn$ may lead
to magnetic frustration in the vicinity of Ag atoms. This scenario
may explain the lack of magnetic signal from the arising $T^*$
anomaly. The same situation occurs for the mirror Ce-triangle
respect to the Ag atom position, placed on the neighboring
Ce-plane. This simultaneous propagation of the AF character
explains why the FM component practically smears out already with
$50\%$ of Ag concentration. On the other hand, the random
fluctuations of magnetically frustrated moments results in a
non-magnetic response of the system.

The low value of $T^*$ and the fact that it does not change with
Ag concentration can be related to $T\to 0$ thermodynamic
constraints. The strong increase of $C_m/T$ at $T>T^*$ cannot be
sustained down to $T=0$ because it would required an available
amount of entropy exceeding the $S_m(T)= \int C_m/T dT = R\ln2$
limit provided by a doublet ground state. To dodge this sort of
{\it entropy bottleneck} \cite{atoms}, the system is driven into
an alternative temperature dependence to allow to reach $S_m=0$
for $T\to 0$, fulfilling the $S_m \leq R\ln2$ condition. In order
to visualize the excess of entropy required by an hypothetical
system whose $C_m/T(T)$ would keep growing below $T^*$ like it
does at $T>T^*$, we have included in Fig.~\ref{F3} a fit performed
on the $C_m/T(T>T^*)$ range as a guide to the eye. Among
alternative functions, we use an heuristic function $C_m/T =
D/(T^Q+E)$ proposed in Ref. \cite{2007} which was already applied
to compare the thermal dependence of some heavy fermion compounds.
Particularly, the entropy associated to this fit nearly doubles
the $R\ln2$ entropy limit.

In summary, we present a FM Ce-system whose $T_C(x)\to 0$, with
the associated degrees of freedom decreasing hand in hand with the
ordering temperature. Such a FM transition extrapolates to a
$T_C=0$ critical point slightly beyond $x = 0.5$. In this system
the magnetic degrees of freedom are progressively transferred to a
{\it non-magnetic} component that emerges as an anomaly centered
at $T^*\approx 1$\,K. The lack of magnetic signal from this
component is attributed to an AF-frustration character driven by
Ce-neighboring Ag atoms, with electron-like character, that modify
the RKKY interaction sign. The constant value of $T^* \approx
1$\,K is attributed to an {\it entropy bottleneck} produced by the
strong increase of magnetic excitations density ($C_m/T$) which
would exceed the available degrees of freedom. This situation
compels $S_m(T)$ to search for a thermodynamic trajectory that
allows to reach $S_m=0$ as $T\to 0$. The microscopic nature of its
ground state remains an open question that requires $\mu$SR
spectroscopy or neutron scattering investigations.


\begin{thebibliography}{00}

\bibitem{Muramatsu} T. Muramatsu, T. Kanemasa, T. Kagayama,
K. Shimizu, Y. Aoki, H. Sato, M. Giovannini, P. Bonville,
V. Zlatic, I. Aviani, R. Khasanov, C. Rusu, A. Amato, K. Mydeen,
M. Nicklas, H. Michor, E. Bauer, Phys. Rev. B 18 (2011) 180404(R).

\bibitem{Fourgeot96} F. Fourgeot, P. Gravereau, B. Chevalier, L. Fournès, J.
Etourneau; J. Alloys and Comp. {\bf 238} 102 (1996).

\bibitem{Mauro00} M .Giovannini, H. Michor, E. Bauer, G. Hilscher, P. Rogl, T. Bonelli, F. Fauth, P. Fischer,
T. Herrmannsdorfer, L. Keller, W. Sikora, A. Saccone, R. Ferro;
Phys. Rev. B {\bf 61} 4044 (2000).

\bibitem{SoliSolut} J.G. Sereni, M. Giovannini, M. G\'omez
Berisso, A. Saccone, Phys. Rev. B {\bf 83} 064419 (2011).

\bibitem{Ramirez} A.P. Ramirez; Annu. Rev. Mater. Sci. {\bf 24} 453
(1994).

\bibitem{SCES_PdRhIn} J.G. Sereni, M. Giovannini, M. G\'omez Berisso,
A. Saccone, J. of Phys.: Conf. Series {\bf 391} 012062 (2012).

\bibitem{Morphol} See for example: P.H. Meijer, J.H. Colwell, B.P. Shah, Am. Jour.
Phys., 41 (1973) 332.

\bibitem{Mydosh} J.A. Mydosh in {\it Spin Glasses: an experimental introduction}.
Taylor \& Francis, London, 1993.

\bibitem{Tari} A. Tari, in {\it The Specific Heat of Matter at low
temperatures}, Imperial College Press, London, Great Britain,
2003.

\bibitem{atoms} J.G. Sereni, J. Low Temp. Phys. {\bf 179} 126 (2015).

\bibitem{2007} J.G. Sereni, J. Low Temp. Phys. {\bf 147} 179
(2007).

\end{thebibliography}
\end{document}